\documentclass[prd,english,preprintnumbers,amsmath,amssymb,nofootinbib,twocolumn,superscriptaddress]{revtex4-1}
\usepackage[utf8]{inputenc}
\usepackage[T1]{fontenc}
\usepackage{amsmath}
\usepackage{amssymb}
\usepackage{graphicx}
\usepackage{tabularx}
\usepackage{bm}
\usepackage{pifont}
\usepackage{mathtools}
\usepackage{graphicx}
\usepackage{xcolor}
\usepackage{graphicx}
\usepackage{subfig}
\usepackage{comment}

\newcommand{\e}{\begin{equation*}\begin{aligned}}
\newcommand{\ee}{\end{aligned}\end{equation*}}

\newcommand{\en}{\begin{equation}\begin{aligned}}
\newcommand{\een}{\end{aligned} \end{equation}}

\newcommand{\bea}{\begin{eqnarray}}
\newcommand{\eea}{\end{eqnarray}}

\newcommand{\p}{\partial}

\newcommand{\f}[2]{\frac{#1}{#2}}

\newcommand{\ra}{\rangle}
\newcommand{\la}{\langle}

\newcommand{\da}{\dagger}
\newcommand{\ma}{\mathcal}

\newcommand{\Q}{\left}
\newcommand{\W}{\right}

\newcommand{\pma}{\begin{pmatrix}}
\newcommand{\epma}{\end{pmatrix}}

\newcommand{\ep}{\epsilon}

\begin{document}

\title{A quantum field-theoretical perspective on scale anomalies in 1D systems \\ with three-body interactions.}
\author{W. S. Daza}
\affiliation{Physics Department, University of Houston. Houston, Texas 77024-5005, USA}
\author{J. E. Drut}
\affiliation{Department of Physics and Astronomy, University of North Carolina, Chapel Hill, North Carolina 27599, USA}
\author{C. L. Lin}
\affiliation{Physics Department, University of Houston. Houston, Texas 77024-5005, USA}
\author{C. R. Ord\'o\~nez}
\affiliation{Physics Department, University of Houston. Houston, Texas 77024-5005, USA}
\affiliation{ICAB,  Universidad de Panam\'a, Panam\'a, Rep\'ublica de Panam\'a.}

\begin{abstract}
We analyze, from a canonical quantum field theory perspective, the problem of one-dimensional particles with three-body attractive interactions, which was recently shown to exhibit a scale anomaly identical to that observed in two-dimensional systems with two-body interactions. We study in detail the properties of the scattering amplitude including both bound and scattering states, using cutoff and dimensional regularization, and clarify the connection between the scale anomaly derived from thermodynamics to the non-vanishing nonrelativistic trace of the energy-momentum tensor.
\end{abstract}

\date{\today}

\maketitle 


\section{Introduction}

The existence of scaling anomalies~\cite{jackiwScaleIntro,thorn,PhysRevD.48.5940,BASUMALLICK2003437} in low-dimensional nonrelativistic systems and its consequences 
in the understanding of ultracold atoms~\cite{RevModPhys.80.885, RevModPhys.80.1215, RevModPhys.82.1225} has recently become the subject of intense activity, both 
theoretically (see e.g.~\cite{PhysRevA.55.R853,PhysRevLett.105.095302, PhysRevA.77.063613, PhysRevA.88.043636, PhysRevLett.106.110403, PhysRevA.92.033603, 
PhysRevLett.115.115301, PhysRevA.93.033639, PhysRevLett.109.135301, PhysRevLett.110.089904, Rev2DLevinsenParish, PhysRevA.97.011602, PhysRevA.97.061603, PhysRevA.97.061604, PhysRevA.97.061605,1807.07106})
and experimentally (see e.g.~\cite{PhysRevLett.105.030404,PhysRevLett.106.105301,PhysRevLett.108.235302,1367-2630-13-11-113032,PhysRevLett.108.070404,PhysRevLett.109.130403,PhysRevLett.112.045301,PhysRevLett.114.110403,PhysRevLett.114.230401,PhysRevLett.115.010401,PhysRevLett.116.045302,PhysRevLett.121.120401,PhysRevLett.121.120402,1805.04734,cite-key}). 
An understanding of the virial expansion entirely within the framework of scaling anomalies was developed for two-dimensional (2D) and one-dimensional (1D) Fermi systems in Refs.~\cite{The2dPaper, drut}, respectively (in the 1D case, three different ``flavors'' of fermions were considered). In 2D, the calculation was based on a quantum field theory (QFT) path-integral representation of the partition function with a two-body local interaction, whereas for the 1D case, a judicious mapping between the quantum-mechanical 2D two-body problem and the quantum-mechanical 1D three-body problem allowed for the treatment of certain aspects of the thermodynamics and the virial expansion of the 1D case [in particular the proportionality between the (interaction-induced) change in the third virial coefficient $\Delta b_3$ in 1D and the
change in the second coefficient $\Delta b_2$ in 2D]. In spite of those advances, a full-fledged QFT treatment of the partition function for the 1D three-body local interaction case is still lacking. 

In this paper, we address the existence of the bound state for the 1D system using canonical QFT methods at zero temperature. Several other issues on anomalies that were addressed for the 2D system in Ref.~\cite{bergmann} are also discussed - we follow this reference closely. Section~\ref{Sec:FirstQ} briefly reviews the quantum-mechanical mapping between the 2D and 1D systems (two-body and three-body respectively), including the bound-state as well as the scattering sector. Section~\ref{Sec:FewBody} states some well-known aspects of nonrelativistic 1D QFT. In Sec.~\ref{Sec:3to3} a calculation of the exact $3 \rightarrow 3$ scattering amplitude is performed; the pole of the amplitude allows one to display the trimer bound-state energy, as well as the running of the dimensionless coupling constant of the three-body local interaction, and the necessary renormalization is made explicit (dimensional transmutation) using a cutoff method. In the following section the same calculation is performed using dimensional regularization (DR).  In Sec.~\ref{Sec:TraceAnomaly}, DR is used to derive the nonrelativistic trace anomaly (dilation anomaly) for the 1D (three-body) and the 2D (two-body) cases. Conclusions and comments are presented in 
Sec.~\ref{section7}. (The number of systems in which a closed-form result of an infinite sum of graphs  for the scattering amplitude is known is rather small.  Additionally, some of the previous derivations of the nonrelativistic trace anomaly in 2D using cutoff methods were not completely satisfactory; the use of DR in this work made the derivation more rigurous -- we do 1D here, but the extension to 2D is straightforward. Hence the pedagogical emphasis of this paper).

\section{First Quantization\label{Sec:FirstQ}}

The 1D three-body Schr\"odinger equation for our system takes the form
\en \label{Eqn:systemEquation}
\left[\frac{-1}{2m}\Q(\frac{\partial^2}{\partial x_1^2}\!+\! \frac{\partial^2}{\partial x_2^2} \!+\! \frac{\partial^2}{\partial x_3^2} \W)+ 
g \delta(x_2 \!-\! x_1)\delta(x_3 \!-\! x_2)\right]
\psi
=
E \psi,
\een
where $\psi = \psi(x_1,x_2,x_3)$ is the 3-body wavefunction. Performing the change of variables
$Q = \frac{1}{3} (x_1+x_2+x_3)$,  $q_1 = x_2  - x_1$, and $q_2 = \frac{1}{{\sqrt{3}}} (x_1 + x_2 - 2 x_3)$, the center of mass (COM) 
factors out and we obtain the relative equation
\en \label{Eqn:2dDelta}
\left[\frac{-1}{2 \tilde{m}}\Q(  \frac{\partial^2}{\partial q_1^2} + \frac{\partial^2}{\partial q_2^2}\W)  + \tilde{g} \, \delta(q_1)\delta(q_2)\right]]\psi = 
E_r \,\psi,
\een
where now $\psi = \psi(q_1, q_2)$, $\tilde m = m/2$, and $\tilde{g} = (2/\sqrt{3})g$. We can treat Eq.~\eqref{Eqn:2dDelta} as a 2D problem for a single particle with mass 
$m/2$. That 2D problem is easily solved, with the result that the system possesses a bound state with energy~\cite{drut,hoffman}
\en\label{Eqn:bareboundJackiw}
E_b=-\f{\Lambda_\text{2D}^2}{2\tilde m}e^{\f{2\pi}{\tilde{m}\tilde{g}}},
\een
where $\Lambda_\text{2D}$ is a cutoff required to make the 2D problem finite. 
After renormalization, the cutoff can be traded for a scale $\mu$ such that
\en\label{Eqn:boundJackiw}
E_b=-\f{\mu^2}{2\tilde m}e^{\f{2\pi}{\tilde{m}\tilde{g}}},
\een
where the coupling $\tilde{g}=\tilde{g}(\mu)$ runs with $\mu$ such that the right hand side of Eq.~\eqref{Eqn:boundJackiw} is an RG invariant. 
For the rest of this work we will set $m=1$ (as we have already done with $\hbar$), such that Eq.~\eqref{Eqn:boundJackiw} reads
\en \label{Eqn:em1boundJackiw}
E_b=-\mu^2 e^{\f{2\sqrt{3}\pi}{g}}.
\een

The $T$-matrix for the 3-body problem can also be solved for using the scattering solution of the 2D problem. The exact scattering solution to Eq.~\eqref{Eqn:2dDelta}  for a particle with incoming momentum $\vec{k}$ is \cite{JackiwDelta}
\e
\phi_k^{\text{in}}(\vec{p})=(2\pi)^2\delta(\vec{p}-\vec{k})+ \\
&&\!\!\!\!\!\!\!\!\!\!\!\!\!\!\!\!\!\!\!\!\!\!\!\!\!\!\!\!
\f{1}{E_k-E_p+i\epsilon}\Q( \f{1}{\tilde{g}}+\f{\tilde{m}}{2\pi}\ln\Q(\f{-\Lambda_\text{2D}^2}{k^2}\W)
\W)^{-1},
\ee
where upon substituting Eq.~\eqref{Eqn:bareboundJackiw}, one arrives at
\e 
\label{Eqn:jackiwscatteringsolution}
\phi_k^{\text{in}}(\vec{p})=(2\pi)^2\delta(\vec{p}-\vec{k})+\\
&&\!\!\!\!\!\!\!\!\!\!\!\!\!\!\!\!\!\!\!\!\!\!\!\!\!\!\!\!
\f{1}{E_k-E_p+i\epsilon}\Q(\f{\tilde{m}}{2\pi}\ln\Q(\f{2\tilde{m}E_b}{k^2}\W)
\W)^{-1}.
\ee

The $T$-matrix $\hat{T}$ can be extracted from the in-state $|\vec{k} \rangle^\text{in}$ projected onto $|\vec{p}\ra$, the scattering state at $t\rightarrow \infty$, via the Lippmann-Schwinger relation \cite{klein}
\en
|\vec{k} \rangle^\text{in}=|\vec{k}\ra+\f{1}{E_k-\hat{H_0}+i\epsilon}\hat{T}|\vec{k}\ra,
\een
from which comparison with Eq.~\eqref{Eqn:jackiwscatteringsolution} gives
\en
\la \vec{p}|\hat{T}|\vec{k}\ra=\Q(\f{\tilde{m}}{2\pi}\ln\Q(\f{2\tilde{m}E_b}{k^2}\W)\W)^{-1}.
\een

Translating from the 2D problem to the 1D problem by using the mapping between 2D and 1D to take $k_1^2+k_2^2=\f{1}{2}(p_2-p_1)^2+\f{1}{6}(p_1+p_2-2p_3)^2$ along with $p_1+p_2+p_3=0$ in the COM frame, one obtains $k_1^2+k_2^2=\f{p_1^2+p_2^2+p_3^2}{2}=E_{\text{COM}}$ such that 
\en
\la \vec{p}|\hat{T}|\vec{k}\ra=\Q(\f{1}{4\pi}\ln\Q(\f{E_b}{E_{\text{COM}}}\W)\W)^{-1}.
\een
Finally, 
\en
\f{\la \vec{p}|\hat{T}|\vec{k}\ra}{\sqrt{\la \vec{p}|\vec{p}\ra \la \vec{k}|\vec{k}\ra}}=\Q(\f{1}{2\sqrt{3}\pi}\ln\Q(\f{E_b}{E_{\text{COM}}}\W)\W)^{-1},
\een
where box normalization is used with $V_x=\f{\sqrt{3}}{2}V_q$ and $V_x$ set to 1 \footnote{Alternatively, using the definition $\tilde{\psi}(\vec{q})\equiv \psi(\vec{x})$ implies the normalization $\la \tilde{\psi} | \tilde{\psi} \ra =\int dq_1dq_2dQ\, \tilde{\psi}^\da (\vec{q})\tilde{\psi}(\vec{q})=\int dx_1 dx_2 dx_3 \,\Big| \f{\p (q_1,q_2,Q)}{\p(x_1,x_2,x_3)}  \Big | \psi^\da(\vec{x}) \psi(\vec{x})=\Big| \f{\p (q_1,q_2,Q)}{\p(x_1,x_2,x_3)} \Big | \int dx_1 dx_2 dx_3 \,  \psi^\da(\vec{x}) \psi(\vec{x})$, with the Jacobian $\Big| \f{\p (q_1,q_2,Q)}{\p(x_1,x_2,x_3)} \Big|=\f{2}{\sqrt{3}}.$ }. 

\section{Brief Review of Few-Body Field Theory\label{Sec:FewBody}}

\begin{figure}[h]
  \centering
  \subfloat[tadpole\label{fig:tadpole}]{\includegraphics[width=3 cm]{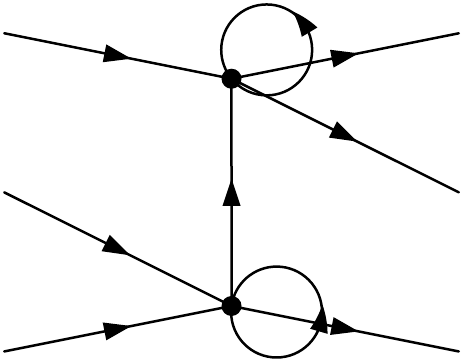}}\hfil
\subfloat[t-channel \label{fig:tchannel}]{\includegraphics[width=3 cm]{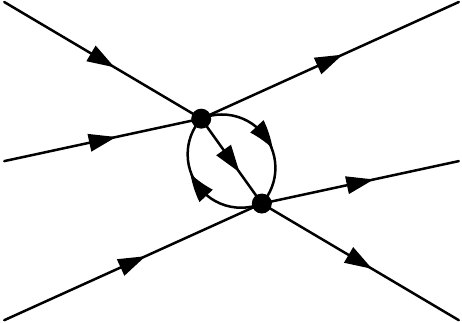}}\hfil
 \subfloat[s-channel \label{fig:schannel}]{\includegraphics[width=3 cm]{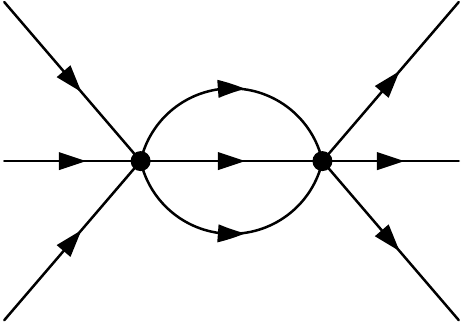}}

    \caption{A few 3 $\rightarrow$ 3, 2-loop diagrams. }\label{fig:oneLoopDiagrams}
\end{figure}

The nonrelativistic Lagrangian density corresponding to Eq.~\eqref{Eqn:systemEquation} is given by \footnote{The label $i$ refers to ``flavor'' quantum number (different spins).}
\en \label{Eqn:lagrangian}
\ma L=\sum_{i=1}^{3}\psi^\da_i\Q(i \p_t + \nabla^2/2 \W) \psi^{}_i-g(\psi^\da_1 \psi^{}_1)(\psi^\da_2 \psi^{}_2)(\psi^\da_3 \psi^{}_3),
\een
such that the free propagator can be read off as $D_{ij}(x,t)=\delta_{ij}D(x,t)$ with
\en
D(x,t) = \int \f{d\omega dk}{(2\pi)^2}\f{e^{-i (\omega t-kx)}}{\omega-k^2/2+i\ep}=-\theta(t)\f{e^{i\Q(\f{x^2}{2t}+\f{\pi}{4}\W)}}{\sqrt{2\pi t}},
\een
which is nonzero only for $t>0$, i.e. propagation forwards in time.\footnote{This is a nonrelativistic few-body ground state, such that holes/antiparticles propagating backwards in time do not exist \cite{Wick}.} We take tadpole graphs $D(0,0)$ to be zero such that the vacuum contains no particles \cite{amateur,boulevard}
\bea
\label{Eqn:tadpole}
&& n(x)=-\lim_{\eta \rightarrow 0^+}\la 0|T\psi(0,x)\psi^\da(\eta,x) |0\ra 
\nonumber \\
&& =-i\lim_{\eta \rightarrow 0^+}\int \f{d\omega dk}{(2\pi)^2}\f{e^{i\eta \omega}}{\omega-k^2/2+i\ep}=0,
\eea
where the contour is completed in the upper half of the complex $\omega$-plane, which misses the pole on the lower half. Alternatively, one can regulate the tadpole with a cutoff, and when quantizing Eq.~\eqref{Eqn:lagrangian}, ambiguity in the ordering of the fields of the interaction term produces a chemical potential counter-term which can cancel the tadpole \cite{bergmann}. Furthermore, if one uses DR, Eq.~\eqref{Eqn:tadpole} is automatically zero \cite{fate}.\\

This feature of propagation only forwards in time makes all diagrams with counterflowing arrows in a loop zero, as $\theta(t_2-t_1)\theta(t_1-t_2)=0$. The vanishing of tadpole and counterflowing graphs implies that only the s-channel graphs, e.g. Fig.~\ref{fig:schannel}, are nonvanishing (see Appendix A).\\


%
\begin{figure}[t]
    \centering
  \subfloat[2 loops\label{fig:wavrebubble}]{\includegraphics[width=3 cm]{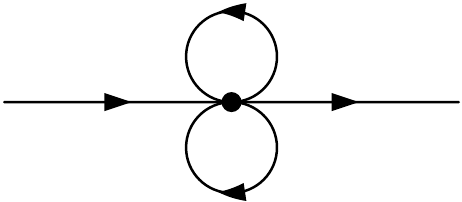}}\hfil
        \subfloat[4 loops\label{fig:fourloopSingPart}]{\includegraphics[width=3 cm]{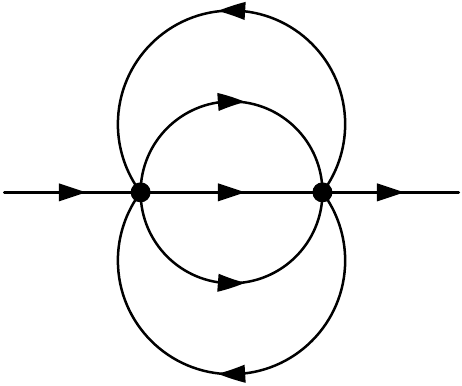}}
    \caption{Lowest order propagator corrections. }\label{fig:propagatorCorrections}
\end{figure}

Moreover, the wavefunction renormalization $Z=1$, as the self energy vanishes such that propagator receives no quantum corrections, 
as illustrated in Fig.~\ref{fig:propagatorCorrections}.

\section{3 $\rightarrow$ 3 Scattering Amplitude\label{Sec:3to3}}

\begin{figure}[b]
    \centering
     \subfloat[tree \label{fig:treeamputate}]{\includegraphics[width=3 cm]{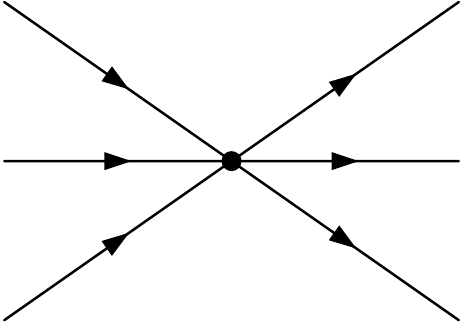}}\hfil
     \subfloat[s-channel \label{fig:oneloopAmputate}]{\includegraphics[width=3 cm]{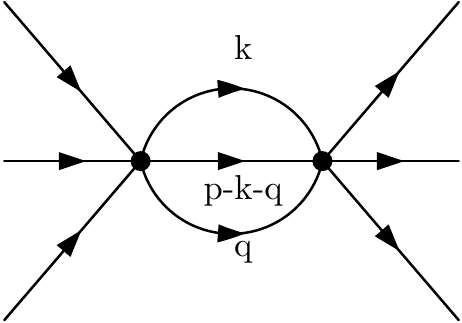}}
    \caption{Lowest order nonvanishing corrections to the $3\rightarrow3$ amplitude. }\label{fig:threeScatter}
\end{figure}

The two-loop contribution to the scattering amplitude is given by Fig.~\ref{fig:oneloopAmputate},
corresponding to the expression
\begin{widetext}
\en \label{13}
A^{(2)}=(-ig)^2\int \frac{d\omega_k d\omega_q dkdq}{(2\pi)^4} \frac{i}{\left(\omega_k-\frac{k^2}{2}+i\epsilon\right)}\frac{i}{\left(\omega_q-\frac{q^2}{2}+i\epsilon\right)}\frac{i}{\left(p_0-\omega_k-\omega_q-\frac{\left(p-k-q\right)^2}{2}+i\epsilon\right)}.
\een
\end{widetext}

\begin{figure}[h]
    \centering
      \includegraphics[scale=1]{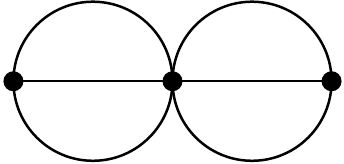}
  \caption{Next term in the geometric series for $3 \rightarrow 3$ scattering.}%
  \label{fig:stringloop}
\end{figure}

The integrals over $\omega_k$ and $\omega_q$ can be done by closing the contour in the lower half complex plane and picking out the poles at $k^2/2-i\epsilon$ and $q^2/2-i\epsilon$, respectively:

\en \label{Eqn:dimReg}
A^{(2)}=\f{-ig^2}{4\pi^2}\int dkdq \frac{1}{p_0-\frac{k^2}{2}-\frac{q^2}{2}-\frac{\left(p-k-q\right)^2}{2}+i\epsilon}.
\een
The remaining momentum integrals can be done successively after completing the square and shifting, using standard integration tables, followed by choosing a cutoff regulator, the result of which is
\en 
A^{(2)}=\f{i g^2}{2\sqrt{3}\pi}\ln\Q(\f{3\Lambda^2}{p^2/6-p_0}\W),
\een
where $A^{(2)}$ is only a function of $p_0-p^2/6$, which is Galilean invariant \footnote{To see this one can go on-shell $p_1^2/2+p_2^2/2+p_3^2/2-(p_1+p_2+p_3)^2/6=\Q((p_1-p_2)^2+(p_2-p_3)^2+(p_3-p_1)^2\W)/6$, and momentum differences are Galilean invariant.}. Adding the tree level-term of Fig. \ref{fig:treeamputate} and stringing together a product of $A^{(2)}$s (see Fig. \ref{fig:stringloop}), one obtains the exact scattering amplitude
\en \label{Eqn:bareAmplitude}
A=\f{-ig_0}{1+\f{g_0}{2\sqrt{3}\pi}\ln\Q(\f{3\Lambda^2}{p^2/6-p_0}\W)},
\een
where the bare coupling $g_0$ was inserted for $g$. 

For a given $g_0$ and $\Lambda$, Eq.~\eqref{Eqn:bareAmplitude} can be replaced with another $g$ and $\mu$ such that $A$ is unchanged for all kinematic values of $p_0-p^2/6$. This is most easily seen by considering the reciprocal of $A$:
\en \label{Eqn:renormalizedEq}
\f{1+\f{1}{2\sqrt{3} \pi}g_0\ln\Q(\f{3\Lambda^2}{p^2/6-p_0}\W)}{-ig_0}&=
\f{1+\f{1}{2\sqrt{3} \pi}g\ln\Q(\f{3\mu^2}{p^2/6-p_0}\W)}{-ig},
\een
which yields
\en
g_0&=\f{g}{1+\f{g}{2\sqrt{3}\pi}\ln\Q(\f{\mu^2}{\Lambda^2}\W)},
\een
where the dependence on the kinematical parameter $p_0-p^2/6$ has dropped out. Therefore in Eq.~\eqref{Eqn:bareAmplitude}, one can always replace bare couplings $(g_0, \Lambda)$ by renormalized ones $(g, \mu)$, so long as they are related by Eq.~\eqref{Eqn:renormalizedEq}. \\

Finally, searching for the pole in Eq.~\eqref{Eqn:bareAmplitude} after going on-shell in the COM frame ($p=\sum\limits_i p_i=0$, $p_0=\sum\limits_i \f{p_i^2}{2}$), followed by continuing into imaginary momenta $p_i\rightarrow i p_i$ gives the trimer bound-state energy
\en\label{Eqn:boundStateEnergyCutoff}
E_b=-3\Lambda^2 e^{\f{2\sqrt{3}\pi }{g_0}},
\een
which can be made to coincide with Eq.~\eqref{Eqn:em1boundJackiw} by using the middle line in Eq.~\eqref{Eqn:renormalizedEq} along with the identification $\sqrt{3}\Lambda=\Lambda_\text{2D}$ (and similarly for $\mu$)
\en
E_b=-\mu^2 e^{\f{2\sqrt{3}\pi }{g}}.
\een
The scattering amplitude $A$ in Eq.~\eqref{Eqn:bareAmplitude} can be written entirely in terms of the bound state energy by using Eq.~\eqref{Eqn:boundStateEnergyCutoff} to get rid of the coupling (dimensional transmutation)
\en \label{Eqn:equation16Amp}
A=\f{-i}{\f{1}{2\sqrt{3}\pi}\ln \Q(\f{E_b}{p_0-p^2/6}\W)}=\f{-i}{\f{1}{2\sqrt{3}\pi}\ln \Q(\f{E_b}{Q_6}\W)}.
\een

\section{Dimensional Regularization\label{Sec:dReg}}

The result of the calculation of the exact scattering amplitude in the previous section  should be independent of the regularization method used. The first calculations in 2D were done using the same cutoff method used here \cite{bergmann}, but dimensional regularization (DR) was also introduced later \cite{fate}. In both cases the system studied was bosonic.  We will now use DR to compute the amplitude, Eq.  (\ref{13}), and then find the exact amplitude.  We will obtain the same result as with momentum cutoff,  showing the robustness of our results. The author of ref. \cite{hoffman} introduced DR to obtain the QFT operator version of the  anomalous current and its non-conservation (Eqs. (\ref{Eqn:r39})-(\ref{44})) for 2D fermions (anomaly equation).  In ref. \cite{bergmann} an expression for the QFT anomaly equation was obtained using a momentum cutoff regularization in a rather indirect fashion.  A direct derivation using DR is particularly suited for this purpose. We present this calculation below for 1D (and as a bonus for 2D).
For this purpose, we denote the number of spatial dimensions as $d$ and write $d=1-\epsilon$, which defines $\epsilon$, and ultimately take 
$\epsilon \rightarrow 0$. In DR, Eq.~\eqref{Eqn:dimReg} holds, but requires two modifications: 1) the bare coupling $g$ (renormalized will be denoted $g_R$) is replaced by $g \mu^{2\epsilon}$, where $g$ is still bare, but dimensionless for arbitrary $\epsilon$, where $\mu$ is an arbitrary scale with units of momentum that absorbs the dimensions of the coupling; 2) the integration measure is $dk^ddq^d$.

To simplify the integral, we use the same trick that we used with cutoff regularization: we note that the final answer can only depend on $p_0-p^2/6$ and therefore we set $p=0$ in the above equation; when we arrive at our final expression, we simply make the replacement $p_0 \rightarrow Q_6 \equiv p_0-p^2/6$ (see Eqs. \eqref{Eqn:dimReg} and \eqref{Eqn:equation16Amp}). In fact, we can make the replacement right away:
\en
A^{(2)}=\f{-ig^2\mu^{4\epsilon}}{(2\pi)^{2d}}\int dk^ddq^d \frac{1}{Q_6-\frac{k^2}{2}-\frac{q^2}{2}-\frac{\left(k+q\right)^2}{2}+i\epsilon}.
\een 
Absorbing $i\epsilon$ into $Q_6$ the denominator can be simplified by rewriting the $k^2+q^2+kq$ as $(k+q/2)^2+3/4 \,q^2$ and making the translational replacement $k+q/2 \rightarrow k$ which has a Jacobian of 1:
\en
A^{(2)}=\f{-ig^2\mu^{4\epsilon}}{(2\pi)^{2d}}\int dk^ddq^d \frac{1}{Q_6-k^2-3q^2/4}.
\een 

Next, we replace $q \rightarrow 2/\sqrt{3} \, q$, which does have a Jacobian:
\en
A^{(2)}=\f{ig^2\mu^{4\epsilon}}{(2\pi)^{2d}} \Q(\f{2}{\sqrt{3}}\W)^{1-\epsilon}\int dk^ddq^d \frac{1}{k^2+q^2-Q_6}.
\een 
Denoting $\ell=(k_1,k_2,...,k_d, q_1,q_2,...,q_d)$ we obtain
\en
A^{(2)}=\f{ig^2\mu^{4\epsilon}}{(2\pi)^{2d}} \Q(\f{2}{\sqrt{3}}\W)^{1-\epsilon}\int d\ell^{2d} \frac{1}{\ell^2-Q_6},
\een 
which is a form appropriate for integration in DR. Note the $i\epsilon$ in $Q_6$ so the integral does not hit a pole. Using Eq.~(8.4) in 
Ref.~\cite{Kleinert:2001ax} with $D \rightarrow 2d$ and $m^2\rightarrow -Q_6$ we obtain

\en
A^{(2)}&=ig^2\mu^{4\epsilon}\Q(\f{2}{\sqrt{3}}\W)^{1-\epsilon}\f{(-Q_6)^{d-1}}{(4\pi)^d}\Gamma(1-d)\\
&=ig^2\mu^{4\epsilon}\Q(\f{2}{\sqrt{3}}\W)^{1-\epsilon}\f{(-Q_6)^{-\epsilon}}{(4\pi)^{1-\epsilon}}\Gamma(\epsilon)\\
&=ig^2\Q(\f{2}{\sqrt{3}}\W)\f{1}{(4\pi)}\mu^{4\epsilon}\Q(\f{2}{\sqrt{3}}\W)^{-\epsilon}\Q(\f{-Q_6}{4\pi}\W)^{-\epsilon}\Gamma(\epsilon)
\een

We use the formula $a+ax+ax^2+...=\f{a}{1-x}$, where $x$ is the ratio of $A^{(2)}$ to $A^{(0)}=-ig \mu^{2\epsilon}$, to obtain the full amplitude:

\en
A =\f{-i\mu^{2\epsilon}}{1/g+\Q(\f{2}{\sqrt{3}}\W)\f{1}{(4\pi)}\mu^{2\epsilon}\Q(\f{2}{\sqrt{3}}\W)^{-\epsilon}\Q(\f{-Q_6}{4\pi}\W)^{-\epsilon}\Gamma(\epsilon)}.
\een

In the denominator we write $\Gamma(\epsilon)=1/\epsilon-\gamma_E $, where $\gamma_E=0.577...$ is the Euler-Mascheroni constant, and note that $a^\epsilon=1+\epsilon \ln a+O[\epsilon^2]$. In addition, we write $1/g=1/g_R^{\overline{\text{MS}}}-\f{1}{2\sqrt{3}\pi}\f{1}{\epsilon}+\f{1}{2\sqrt{3}\pi}\Q(\gamma_E+\ln\Q(\f{2}{\sqrt{3}}\W)-\ln(4\pi)\W)$ for $\overline{\text{MS}}$ \cite{srednicki}. Then as $\epsilon \rightarrow 0$,
\en \label{Eqn:dTrans29}
A&=\f{-i }{1/g^{\overline{\text{MS}}}_R+\f{1}{2\pi \sqrt{3}}\ln\Q(\f{\mu^2_{\overline{\text{MS}}}}{-Q_6}\W)}.
\een
Moreover, from 
\en
\frac{1}{g}=\frac{1}{g_R^{\overline{\text{MS}}}}-\f{1}{2\sqrt{3}\pi}\f{1}{\epsilon}+\f{1}{2\sqrt{3}\pi}\Q(\gamma_E+\ln\Q(\f{2}{\sqrt{3}}\W)-\ln(4\pi)\W),
\een
we see that to order $1/\epsilon$, we can make the replacement
\en \label{Eqn:amount30}
\frac{1}{g} \rightarrow -\f{1}{2\sqrt{3}\pi}\f{1}{\epsilon}=\f{1}{2\sqrt{3}\pi}\f{1}{d-1},
\een
which amounts to $\f{1-d}{g} \rightarrow -\f{1}{2\sqrt{3}\pi} $.\\

Finally, we will implement dimensional transmutation in Eq.~\eqref{Eqn:dTrans29}: in the center-of-mass frame ($p=0$), after analytic continuation $p_0=\sum \limits_{i=1}^3 \f{p_i^2}{2} \rightarrow -\sum \limits_{i=1}^3 \f{p_i^2}{2}$, we find the bound state energy $E_b=-\sum \limits_{i=1}^3 \f{p_i^2}{2}$ from the pole of Eq.~\eqref{Eqn:dTrans29}
\en
\frac{1}{g^{\overline{\text{MS}}}_R}+\f{1}{2\pi \sqrt{3}}\ln\Q(\f{\mu^2_{\overline{\text{MS}}}}{-E_b}\W)=0.
\een

Therefore,
\begin{widetext}
\en
1/g^{\overline{\text{MS}}}_R+\f{1}{2\pi \sqrt{3}}\ln\Q(\f{\mu^2_{\overline{\text{MS}}}}{-Q_6}\W)&=1/g^{\overline{\text{MS}}}_R+\f{1}{2\pi \sqrt{3}}\ln\Q(\f{\mu^2_{\overline{\text{MS}}}}{-Q_6}\W)+\Q(\f{1}{2\pi \sqrt{3}}\ln\Q(\f{-E_b}{\mu^2_{\overline{\text{MS}}}}\W)-\f{1}{2\pi \sqrt{3}}\ln\Q(\f{-E_b}{\mu^2_{\overline{\text{MS}}}}\W)\W)\\
&=1/g^{\overline{\text{MS}}}_R+\f{1}{2\pi \sqrt{3}}\ln\Q(\f{\mu^2_{\overline{\text{MS}}}}{-E_b}\W)+\f{1}{2\pi \sqrt{3}}\ln\Q(\f{E_b}{Q_6}\W)\\
&=\f{1}{2\pi \sqrt{3}}\ln\Q(\f{E_b}{Q_6}\W),
\een
\end{widetext}
and Eq.~\eqref{Eqn:dTrans29} becomes
\en
A=\f{-i}{\f{1}{2\pi \sqrt{3}}\ln\Q(\f{E_b}{Q_6}\W)},
\een
which is the same as Eq.~\eqref{Eqn:equation16Amp}.

\section{Trace Anomaly\label{Sec:TraceAnomaly}}

As is well known, Noether's theorem gives a constructive procedure to find conserved charges, the so-called Noether charges \cite{Weinberg:1995mt}, whenever the classical action for a field theory is invariant under global symmetry transformations. In the particular case of the 1D nonrelativistic Lagrangian, Eq.~\eqref{Eqn:lagrangian},  the following classical conservation equation is obtained:
\en\label{Eqn:consD}
\p_\mu j^\mu=0,
\een
where  $j^\mu=\Q(\f{\p \ma L}{\p \p_\mu\phi_i}\delta\phi_i+f^\mu \ma L\W)$. Equation~\eqref{Eqn:consD} is 
derived in Appendix \ref{acnoeth}. In the definition of $j^\mu$, $\phi_i=(\psi_a,\psi^*_{a}, a=1,2,3),$ 
$\delta \phi_i \rightarrow \delta \psi_a=-(f^\mu \p_\mu+d/2)\psi_a$, $f^\mu=(2\tau,\vec{x}),$ and similarly for $ \delta \psi_a^*$. The existence of an anomaly implies that the right-hand side of the  operator version of this equation will not be zero.  We will calculate it here using our previous DR procedure and results. Following Ref.~\cite{hoffman}, we will also show a derivation for the integral version of the anomaly equation.

To use DR, we have to replace the coupling constant $g$ as in Sec.~\ref{Sec:dReg} by $g\mu^{2\epsilon}$, with $\epsilon=1-d$. In this fashion, the interaction term $\ma L_I$  is no longer invariant, only the free part $\ma L_0$ is\footnote{This means that we cannot apply the results of Appendix  \ref{acnoeth} to $\mathcal{L}$, Eq. (\ref{Eqn:lagrangian}), in a straight fashion and had to start with Eq. (\ref{35}) in order to arrive at Eq. (\ref{Eqn:eqn37r}).}. Following the procedure of Appendix \ref{acnoeth}, the finite version of the variation of $\ma L_0$ is 
\en \label{35}
\delta \ma L_0=\f{\p \ma L_0}{\p \phi_i}\delta\phi_i+
\f{\p \ma L_0}{\p \p_\mu \phi_i}\p_\mu (\delta\phi)_i
=-\p_\mu(\ma L_0 f^\mu).
\een

Since $\f{\p \ma L_I}{\p (\p_\mu \phi_i)}=0$, we can use $\ma L_0=\ma L-\ma L_I$ and the equation of motion $\f{\p \ma L}{\p \phi_i}=\p_\mu \Q(\f{\p \ma L}{\p (\p_\mu \phi_i)}\W)$ to obtain
\en\label{Eqn:36r}
&\p_\mu \Q(\f{\p \ma L}{\p (\p_\mu \phi_i)}\W)\delta \phi_i+
\f{\p \ma L}{\p (\p_\mu \phi_i)}\p_\mu\Q(\delta \phi_i\W) \\
&= -\p_\mu\Q(\ma L f^\mu \W)+\p_\mu\Q(\ma L_I f^\mu\W)+\f{\p \ma L_I}{\p  \phi_i}\delta\phi_i.
\een

Identifying $j^\mu\equiv \Q(\f{\p \ma L}{\p \p_\mu\phi_i}\delta\phi_i+ \ma L f^\mu\W)$, we obtain
\bea
j^0 &=& \frac{x}{2 i} [ \psi^\dagger_\sigma \partial_x \psi_\sigma-(\partial_x \psi^\dagger_\sigma )\psi_\sigma  ]-2\tau \mathcal H \\
j^1 &=& x[\mathcal L+\partial_x \psi^\dagger_\sigma \partial_x \psi_\sigma]+\tau[ \partial_x \psi^\dagger_\sigma \partial_t \psi_\sigma+\partial_t \psi^\dagger_\sigma \partial_x \psi_\sigma] \nonumber  \\
&& + \frac{1}{4}\partial_x(\psi^\dagger_\sigma\psi_\sigma),
\eea
where
%
%
\en
\ma H=-\sum_{i=1}^{3}\psi^\dagger_i\left( \nabla^2/2 \right) \psi^{}_i+g(\psi^\dagger_1 \psi^{}_1)(\psi^\dagger_2 \psi^{}_2)(\psi^\dagger_3 \psi^{}_3).
\een

Equation~\eqref{Eqn:36r} can then be written as\footnote{Here and elsewhere the summation over $i$ means summation over $a$ and both $\psi_a$ and $\psi_a^*$; e.g., $\f{\p \ma L_I}{\p\phi_i}\delta \phi_i\rightarrow\sum\limits_{a=1}^3 \f{\p \ma L_I}{\p\psi_a}\delta \psi_a+\sum\limits_{a=1}^3 \f{\p \ma L_I}{\p\psi^*_a}\delta \psi^*_a$.}
\en\label{Eqn:eqn37r}
\p_\mu j^\mu&=\p_\mu(\ma L_I f^\mu) +\f{\p \ma L_I}{\p \phi_i}\delta\phi_i\\
&=(2+d)\ma L_I-\f{d}{2}\f{\p \ma L_I}{\p \phi_i}\phi_i,
\een

where we used $\p_\mu f^\mu=(d+2)$ and $\delta \phi_i=-\Q(d/2+f^\mu \p_\mu \W)\phi_i$ (see Appendix \ref{acnoeth}). \\

The derivation of Eq.~\eqref{Eqn:eqn37r} also applies for the case of $d=2$ with $\ma L=\ma L_0-g\Q(\psi^*\psi \W)^2$ \cite{bergmann,PhysRevD.49.4299} (and its fermionic version \cite{The2dPaper}). In this case one can easily see that $\p_\mu j^\mu \propto (d-2)$, so classically $\p_\mu j^\mu=0$. Likewise, for the Lagrangian \eqref{Eqn:lagrangian} we find
\en
\p_\mu j^\mu=-2(1-d)g\mu^{2\epsilon}(\psi^*_1\psi^{}_1)(\psi^*_2\psi^{}_2)(\psi^*_3\psi^{}_3).
\een

Again, classically $d\rightarrow 1$ ($\epsilon\rightarrow 0$; no running of $g$) gives $\p_\mu j^\mu=0$. However, quantum-mechanically, $\f{1-d}{g} \rightarrow -\f{1}{2\pi\sqrt{3}}$ as shown in Sec.~\ref{Sec:dReg}; therefore, for $d \rightarrow 1$, we obtain the trace (dilation) anomaly equation
\en \label{Eqn:r39}
\p_\mu j^\mu=\f{1}{\sqrt{3}\pi}g^2(\psi^\da_1\psi^{}_1)(\psi^\da_2\psi^{}_2)(\psi^\da_3\psi^{}_3).
\een

Using the results of \cite{hagen1} we obtain\footnote{Notice $g$ is running, $g\propto(d-1)$ (Eq.~\eqref{Eqn:amount30}). As in the 2D case, the matrix elements of the operator on the RHS of Eq.~\eqref{Eqn:r39} are expected to diverge as $(d-1)^{-1}$, rendering the matrix elements of Eq.~\eqref{Eqn:r40} finite \cite{PhysRevLett.100.205301}.}
\en \label{Eqn:r40}
2\hat{h}-\hat{T}_{xx}=-\f{1}{\sqrt{3}\pi}g^2(\psi^\da_1\psi^{}_1)(\psi^\da_2\psi^{}_2)(\psi^\da_3\psi^{}_3).
\een
where $\hat{T}_{ij}=$ is the energy-momentum tensor and $\hat{h}=$ energy density; see below. Equation~(\ref{Eqn:r40}) is the 1D analogue of the 2D version 
\en \label{44}
2\hat{h}-\sum\limits_{i=1}^2\hat{T}_{ii}=-\f{1}{2\pi}g^2 \psi^\da_\uparrow \psi^\da_\downarrow\psi^{}_\downarrow \psi^{}_\uparrow,
\een
which can be derived using this method; see also \cite{hoffman}. Notice we have now replaced the classical fields $\psi^*_a$, $\psi^{}_a$ by operators $\psi^\da_a$, $\psi^{}_a$. The RHS of Eq.~\eqref{Eqn:r40} is the three-body analog of the Tan contact density from two-body physics \cite{The2dPaper, hoffman,TAN20082952,TAN20082971,TAN20082987}. This identification is made more clear following Hofmann's derivation for the 2D case, which we now describe. \\

In order to use DR we need the expression for the $d$-dimensional energy-momentum tensor. The energy-momentum tensor of the three-body zero range model in $d$-dimensions is \footnote{In what follows we will omit the $\mu^{2\epsilon}$ since it will not contribute in the $\epsilon\rightarrow0$ limit.} 
\begin{widetext}
\en 
\hat{T}_{ij}=\frac{1}{2}\left(\p_i\psi^\dagger_\sigma\p_j\psi^{}_\sigma+\p_j\psi^\dagger_\sigma\p_i\psi^{}_\sigma-\frac{\delta_{ij}}{2}\nabla^2\left(\psi^\dagger_\sigma\psi^{}_\sigma\right)\right)+2g\delta_{ij}\left(\psi^\dagger_1\psi^{}_1\right)\left(\psi^\dagger_2\psi^{}_2\right)\left(\psi^\dagger_3\psi^{}_3\right),\\
\een
\end{widetext}
where $\sigma=1,2,3.$ One can take the $d-$dimensional trace
\en 
\hat{T}_{ii}&=2\left(\frac{1}{2}\p_i\psi^\dagger_\sigma\p_i\psi^{}_\sigma+g\left(\psi^\dagger_1\psi^{}_1\right)\left(\psi^\dagger_2\psi^{}_2\right)\left(\psi^\dagger_3\psi^{}_3\right)\right)\\
&\ \ +2\frac{(d-1)}{g}g^2\left(\psi^\dagger_1\psi^{}_1\right)\left(\psi^\dagger_2\psi^{}_2\right)\left(\psi^\dagger_3\psi^{}_3\right)
\\
&\ \ -\frac{d}{4}\nabla^2\left(\psi^\dagger_\sigma\psi^{}_\sigma\right).
\een

The first term on the RHS is twice the Hamiltonian density, the second term
naively vanishes when $d = 1$, and the last term vanishes upon integration over
space. Performing such an integral,
\en
\int d^dx \, \hat{T}_{ii}=2\hat{H}+2\frac{(d-1)}{g}I
\een
where we define the 3-body contact,
\en
I\equiv g^2\int d^dx \left(\psi^\dagger_1\psi^{}_1\right)\left(\psi^\dagger_2\psi^{}_2\right)\left(\psi^\dagger_3\psi^{}_3\right),
\een
and using Eq.~(\ref{Eqn:amount30}) we obtain
\en
2\hat{H}-\int dx\, \hat{T} _{xx}=-\frac{1}{\sqrt{3}\pi}I.
\een

\section{Conclusions}\label{section7}

In this paper we used perturbative methods of quantum field theory to study the bound-state and scattering problem of three different species of fermions interacting via three-body local interactions in 1D. The quantum mechanical version (first quantization) and some aspects of thermodynamics were studied in Ref.~\cite{drut}. A necessary summation to all orders in the $3\rightarrow 3$ scattering matrix was performed, using both cutoff and dimensional regularization. We confirmed the results of \cite{drut} that were based on a judicious change of variables that related the 1D to the 2D similar problem. A derivation of the trace (dilation) anomaly in 1D and 2D was given using the DR results developed in this paper. An interesting question is how far can one go with perturbative methods in understanding the many-body behavior of the system, where there is both anomalous breaking of scale symmetry as well as spontaneous symmetry breaking. In our previous work we explored the many-body behavior of the system using a lattice approach~\cite{drut}. While the simplicity of few-body allows us to get exact results with perturbative methods, we anticipate that a perturbative calculation of the many-body behavior will be limited to the first fewest loops.  We leave this for future work.
\section*{Acknowledgements}\label{section8}
This work was supported
by the U.S. National Science Foundation under Grant No.
PHY1452635 (Computational Physics Program). This
work was supported in part by the U.S. Army Research
Office Grant No. W911NF-15-1-0445.


\appendix

\section{Vanishing of non-S-channel Graphs}

\begin{figure}[h]
    \centering
      \includegraphics[scale=1]{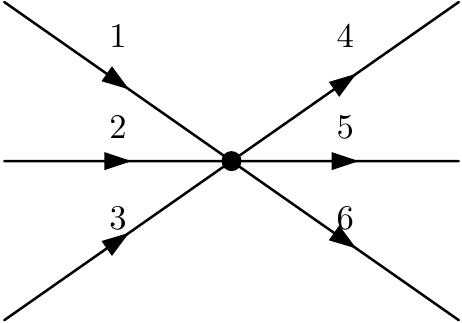}
    ~ 
    ~ 

  \caption{3-body vertex.}%
  \label{fig:loopVanish}
\end{figure}

We proceed to show using Fig. \ref{fig:loopVanish} that all internal lines leaving or entering a vertex that participate in loops must have arrows in the same direction, or else the diagram vanishes. For the sake of argument, suppose 3, 5, and 6 participate in loops. Denoting $p_0$ and $p$ as the sum of the frequencies and momenta of 1, 2, and 4, respectively, then the loop integrals are proportional to
\begin{widetext}
\en
\int d\omega_5 d \omega_6 \f{1}{\omega_5-k_5^2/2+i\epsilon} \f{1}{\omega_6-k_6^2/2+i\epsilon}\f{1}{(\omega_5+\omega_6-p_0)-(k_5+k_6-p_0)^2/2+i\epsilon},
\een
\end{widetext}
which vanishes when completing the contour in the upper half of the complex frequency plane as all poles are below the real axis. Note that it is not critical that both 5 and 6 participate in loops: we only need one of the internal lines to participate in a loop in order to force the remaining internal lines to be in the same direction.

\section{Symmetries of the Action and Noether's current Equation}\label{acnoeth}
Consider the action 
\[
S[\phi_i,V,T]=\int_{VT} dx \, \mathcal{L}(\phi_i(x),\p_\mu\phi_i(x)),
\]
where the only spacetime dependence will be through the fields $\phi_i(x)$, $dx$ represents the spacetime measure $dx\equiv d^dx dt$ ($VT$ is the spacetime volume; $V$ usually taken to be very large). The action is to be symmetric under the simultaneous transformation of the coordinates and the fields ($R_{ij}$ will be taken to be spacetime independent)
\en 
{x}'^\mu&={x'}^\mu (x^\nu)\\
\phi'_i(x')&=R_{ij}\phi_j(x) \\
\phi '(x)&=R_{ij}\phi_j(x^{-1}),
\een
such that
\en
S[\phi_i',V',T']=S[\phi_i,V,T],
 \een 
 \en 
 \int_{V'T'}dx'\mathcal{L}(\phi'_i(x'),\p_\mu'\phi'_i(x'))=\int_{VT} dx \,\mathcal{L}(\phi_i(x),\p_\mu\phi_i(x)).
 \een

Going from $V'T'$ to $VT$ will produce a Jacobian
  \en 
 &\int_{VT}dx\left|\frac{\p x'}{\p x}\right|\mathcal{L}(R_{ij}\phi_j(x),\left(\frac{\p x^\nu}{\p x'^\mu}\right)\p_\nu R_{ij}\phi_j(x))\nonumber \\
 &\quad=\int_{VT} dx \mathcal{L}(\phi_i(x),\p_\mu\phi_i(x)),
 \een
 or simply
  \en 
\left|\frac{\p x'}{\p x}\right|\mathcal{L}(R_{ij}\phi_j(x),\left(\frac{\p x^\nu}{\p x'^\mu}\right)\p_\nu R_{ij}\phi_j(x))= \mathcal{L}(\phi_i(x),\p_\mu\phi_i(x))\label{lagact}.
 \een
 
Let us consider an infinitesimal transformation, $R_{ij}=\delta_{ij}+\eta r_{ij}$,
such that
\en 
{x}'^\mu&={x}^\mu+\eta f^\mu(x^\nu),
\een
where $f^\mu=(2t,\vec{x})$, $r_{ij}=-\f{d}{2}\delta_{ij}$, and so
\en
\phi'_i(x')&=\phi_i(x)+\eta r_{ij}\phi_j(x) \\
\phi '(x)&=\phi_i(x)+\eta r_{ij}\phi_j(x)-\eta f^\nu\p_\nu \phi_i(x).
\een
Then, 
\en 
\frac{\p x'^\mu}{\p x^\nu}=\delta^\mu_\nu +\eta \left(\p_\nu f^\mu\right), 
\een 
such that
\en
\left|\frac{\p x'}{\p x}\right|=1+\eta \p_\nu f^\nu+O(\eta^2), 
\een
and
\bea
R_{ij}\phi_j(x)&=&\left(\delta _{ij}+\eta r_{ij}\right)\phi_j(x) \nonumber \\
&=&\phi_i(x)+\eta \left(\delta \phi _i(x)+f^\nu\p_\nu \phi_i(x)\right),
\eea 
where we have defined $\eta\delta\phi_i(x)=\phi_i'(x)-\phi_i(x)$ such that
\en 
\delta \phi_i(x)=r_{ij}\phi_j(x)-f^\nu\p_\nu \phi_i(x),
\een 
and
\en 
\delta \p_\mu\phi_i(x)=\p_\mu\delta\phi_i(x)=r_{ij}\p_\mu \phi_j(x)-\p_\mu\left(f^\nu\p_\nu \phi_i(x)\right).
\een 

When the above is replaced into Eq.~(\ref{lagact}) to order $\eta$ we obtain
\begin{widetext}
\en 
\left|\frac{\p x'}{\p x}\right|\mathcal{L}(\phi_i(x)+\eta \delta \phi _i(x)+\eta f^\nu\p_\nu \phi_i(x),\p_\mu\left(\phi_i(x)+\eta \delta \phi _i(x)+\eta f^\nu\p_\nu \phi_i(x)\right)-\eta \left(\p_\mu f^\nu \right)\p_\nu\phi_i(x)) = \mathcal{L}(\phi_i(x),\p_\mu\phi_i(x)).
\een 

Performing a Taylor expansion and keeping only terms to order $\eta$ gives
\en 
0=\eta \p_\nu f^\nu \mathcal{L}+\eta \frac{\p\mathcal{L}}{\p\phi_i(x)}\left[\delta\phi_i(x)+f^\nu\p_\nu\phi_i(x)\right]+\eta\frac{\p\mathcal{L}}{\p\p_\mu\phi_i(x)}\left[\p_\mu\delta\phi_i(x)+f^\nu\p_\nu\p_\mu\phi_i(x)\right],\\\
\een 
\end{widetext}

leading to the desired expression
\en 
\delta \mathcal{L}\equiv \frac{\p\mathcal{L}}{\p\phi_i(x)}\delta \phi_i(x)+ \frac{\p\mathcal{L}}{\p\p_\mu\phi_i(x)}\p_\mu\delta\phi_i(x)=-\p_\nu\left(f^\nu\mathcal{L}\right).
\een

Using the equation of motion $\frac{\p\mathcal{L}}{\p\phi_i(x)}=\p_\mu\frac{\p\mathcal{L}}{\p\p_\mu\phi_i(x)}$, this leads to Noether's current equation
\en 
\p_\mu j^\mu=\p_\mu\left[\frac{\p\mathcal{L}}{\p\p_\mu\phi_i(x)}+f^\mu\mathcal{L}\right]=0.
\een 

\end{document}